\documentclass[prd,twocolumn,nofootinbib,showpacs,superscriptaddress]{revtex4}

\usepackage{amsmath}
\usepackage{latexsym}
\usepackage{graphicx}
\usepackage{color}

\include{macros}
\newcommand{\etal}{{\it et al.~}\!}

\begin{document}

\title{Recoiling from a kick in the head-on collision of spinning black holes}

\author{Dae-Il Choi}
\affiliation{Gravitational Astrophysics Laboratory, 
NASA Goddard Space Flight Center, 8800 Greenbelt Rd., 
Greenbelt, MD 20771, USA}
\affiliation{Korea Institute of Science and Technology Information, \\
52-11, Eoun-Dong, Yuseong-Gu, Daejeon, South Korea, 305-806}
\author{Bernard J. Kelly}
\affiliation{Gravitational Astrophysics Laboratory,
NASA Goddard Space Flight Center, 8800 Greenbelt Rd.,
Greenbelt, MD 20771, USA}
\author{William D. Boggs}
\affiliation{University of Maryland, Department of Physics, College Park, MD 20742, USA}
\author{John G. Baker}
\affiliation{Gravitational Astrophysics Laboratory, 
NASA Goddard Space Flight Center, 8800 Greenbelt Rd.,
Greenbelt, MD 20771, USA}
\author{Joan Centrella}
\affiliation{Gravitational Astrophysics Laboratory, 
NASA Goddard Space Flight Center, 8800 Greenbelt Rd., 
Greenbelt, MD 20771, USA}
\author{James van Meter}
\affiliation{Gravitational Astrophysics Laboratory, 
NASA Goddard Space Flight Center, 8800 Greenbelt Rd., 
Greenbelt, MD 20771, USA}

\date{\today}

\begin{abstract}
Recoil ``kicks'' induced by gravitational radiation are expected in
the inspiral and merger of black holes. Recently the numerical
relativity community has begun to measure the significant kicks found
when both unequal masses and spins are considered. Because
understanding the cause and magnitude of each component of this kick
may be complicated in inspiral simulations, we consider these effects
in the context of a simple test problem.  We study recoils from
collisions of binaries with initially head-on trajectories, starting
with the simplest case of equal masses with no spin and then adding
spin and varying the mass ratio, both separately and jointly. We find
spin-induced recoils to be significant relative to unequal-mass
recoils even in head-on configurations. Additionally, it appears that
the scaling of transverse kicks with spins is consistent with
post-Newtonian theory, even though the kick is generated in the
nonlinear merger interaction, where post-Newtonian theory should not
apply. This suggests that a simple heuristic description might be
effective in the estimation of spin-kicks.
\end{abstract}


\maketitle

\section{Introduction}
\label{sec:introduction}

The coalescence of spinning black holes in a binary system is expected
to occur throughout the universe, on scales ranging from stellar black
holes formed as the end-products of stellar evolution to supermassive
black holes that lurk at the centers of galaxies.  The final merger of
such systems will produce an intense burst of gravitational radiation;
if this radiation is emitted asymmetrically, as in the case of unequal
masses and spins, the resulting remnant black hole will experience a
recoil kick.  The magnitude of this kick is important in a variety of
astrophysical situations, such as the cosmological evolution of
supermassive black holes \cite{Merritt:2004xa,Boylan-Kolchin:2004tf,
Haiman:2004ve,Madau:2004st,Yoo:2004ze,Volonteri:2005pn,
Libeskind:2005eh,Micic:2006ta} and the growth and retention of
intermediate-mass black holes in dense stellar clusters
\cite{Miller:2001ez,Miller:2002pg,Mouri:2002mw,Miller:2003sc,
Gultekin:2004pm,Gultekin:2005fd,O'Leary:2005tb}, and it also
affects the expected rates of black hole mergers for gravitational
wave detectors \cite{Sesana:2004gf}.
Given the importance of recoil kicks in astrophysics, there have
been numerous analytic studies of this phenomenon
\cite{Peres62,Bekenstein1973,Fitchett_1983,
FD1984,RR89,Wiseman92,Favata:2004wz,Blanchet:2005rj,Damour:2006tr}.
However, since nearly all of the recoil occurs in the regime of strong
gravitational fields, numerical relativity simulations are essential
to obtain accurate calculations of the kick.

Recent breakthroughs in binary black hole simulations
\cite{Pretorius:2005gq,Campanelli:2005dd,Baker:2005vv,
Campanelli:2006gf} now allow extensive studies of equal mass
nonspinning black hole mergers
\cite{Baker:2006yw,Buonanno:2006ui,Baker:2006kr}. 
Calculations of the kick resulting from mergers of nonspinning
black holes with unequal masses have been carried out for
mass ratios $q = m_1/m_2$ in the range $q = 1$ to $q = 0.25$
\cite{Herrmann:2006ks,Baker:2006vn,Gonzalez:2006md}, 
with the maximum kick velocity estimated to be
$\sim 175 {\rm km/s}$ for mass ratio $q \sim 0.36$
\cite{Gonzalez:2006md}.  In addition, several simulations of
mergers of equal mass, spinning black holes have been carried
out \cite{Campanelli:2006uy,Campanelli:2006fg,Campanelli:2006fy}.

Most recently, several papers have appeared that address the kicks
obtained from the inspiral of spinning binaries of equal
\cite{Herrmann:2007ac,Koppitz:2007ev} and unequal masses
\cite{Baker:2007gi}. These papers find considerable kicks
resulting from the addition of symmetry-breaking spins perpendicular
to the orbital plane, leading to a total kick speed of $\sim 440 {\rm
km/s}$. Moving to spins initially in the orbital plane,
\cite{Campanelli:2007ew} suggested a ``superkick'' orbital
configuration; this was investigated by Gonzalez
\etal \cite{Gonzalez:2007hi}, who found an associated kick of
$\sim 2500 {\rm km/s}$. Extrapolated to maximally spinning pre-merger holes, 
this may lead to a kick as high as $\sim 4000 {\rm km/s}$ \cite{Campanelli:2007ew}.

Nevertheless the complexity of the binary orbital evolution may
obscure details such as the direction of the final kick, and its
dependence on mass ratio and spin. With this in mind, we present here
a study of a simpler problem, which may be seen as an approximation to
the final plunge to merger. We chose the head-on case as a model
problem to isolate kick effects from the orbital inspiral
motion. Although lacking in astrophysical likelihood, our head-on
investigations have the advantage of removing directional ambiguity in
the kicks produced.  In particular, we can readily test the
leading-order post-Newtonian (PN) prediction that the spin and
mass-ratio contributions to the kick should be orthogonal.

In our investigation of the recoil kicks produced by the merger of
spinning black holes in head-on collisions, we vary both the black
hole spins and mass ratio.  We find definite transverse kicks from the
merger of equal-mass holes with spin. Furthermore, the total kick
momentum imparted appears to scale roughly with the sum of the black
hole spin parameters, $a_1+a_2$.  For the cases investigated, these
kicks yield transverse kick velocities in the range $\sim 15 - 30 {\rm
km/s}$. We have also seen longitudinal (along the line of motion of
the holes) kicks, due to a non-unity mass ratio. These kicks are much
more modest in magnitude, with velocities of $\sim (2 - 5) {\rm km/s}$
for the mild mass ratio $q=2/3$ chosen.

This work is carried out using the moving puncture method
\cite{Campanelli:2005dd,Baker:2005vv,
vanMeter:2006vi,Hannam:2006vv,Bruegmann:2006at,Sperhake:2006cy}.  We
describe our initial data in Sec.~\ref{sec:init} and our numerical
methodology in Sec.~\ref{sec:methodology}.  Code calibration and
testing is presented in Sec.~\ref{sec:tests}.  We present our results
in Sec.~\ref{sec:results} and conclude with a discussion in
Sec.~\ref{sec:discussion}.

\section{Initial Data}
\label{sec:init}

We set up initial data for binary black holes represented as
``punctures'' \cite{Brandt97b}.  The metric on the initial spacelike
slice is written in the form $g_{ij} = \psi^4\delta_{ij}$, where $i,j
= 1,2,3$, with conformal factor $\psi = \psi_{\rm BL} + u$. The
singular part of the conformal factor takes the form $\psi_{\rm BL} =
1 + \sum_{A=1}^{2} m_{A,p}/2 |\vec{r} - \vec{r}_A|$, where the $A^{\rm
th}$ black hole has mass parameter (or ``puncture mass'') $m_{A,p}$
and is located at coordinate position $\vec{r}_A$.  The nonsingular
function $u$ is calculated by solving the Hamiltonian constraint
equation using the second-order-convergent elliptic solver {\tt AMRMG}
\cite{Brown:2004ma}.

The black holes start out at rest. We use the Bowen-York (BY)
\cite{Bowen80} form of the extrinsic curvature to incorporate black
hole spin:
\begin{equation}
K^S_{A,ij} = \psi^{-2} \frac{3}{r_A^3} \left[ \epsilon_{kim} S_A^m n_A^k n_{A\,j} + \epsilon_{kjm} S_A^m n_A^k n_{A \, i} \right], \label{eqn:BYspin} 
\end{equation}
where $\vec{n}_A$ is the unit vector in the direction of increasing
$r_A$ and $\vec{S}_A$ is the spin angular momentum of the $A^{th}$
black hole. In all cases we take $\vec{S}_A$ to be aligned with the
$z$ axis, so that $\vec{S}_A = S_A \hat{e}_z$.

In all, we consider head-on collisions for seven different initial
configurations of both equal (EQ) and unequal (NE) masses, and varied
spins. The initial directly specified parameters of all the simulations are listed in Table
\ref{table:idparams}{, while derived quantities are
given in Table~\ref{table:idderived}}; all length scales are given in
terms of a fiducial mass $M$ [which coincides with the
Arnowitt-Deser-Misner (ADM) mass of the system for equal-mass
runs]. For each run, the punctures were initially placed on the $y$
axis with the center of mass at the origin. The proper separation $l$
between the holes, along the initial slice, is measured between the
closest parts of the apparent horizons along the $y$ axis. We note
that $l$ is not necessarily the smallest physical distance between the
holes (in this spatial slice), as spin effects can twist the
space-like geodesics off the coordinate axis.
\begin{center}
\begin{table}[t]
\caption{Directly specified parameters of the numerical simulations, in
terms of the fiducial mass $M$. $S_A^z$ is the non-zero component of
the Bowen-York angular momentum on each hole. ${\rm NEa}_{+-}$ and
${\rm NEb}_{+-}$ are, respectively, large-separation and
equal-spin-parameter variants of ${\rm NE}_{+-}$, as explained in the
text.}
\label{table:idparams}
 \begin{tabular}{c| c c c c c c}
  \hline  \hline
   Run      & $m_{1,p}/M$ & $m_{2,p}/M$ & $y_1/M$ & $y_2/M$ & $S^z_1/M^2$ & $S^z_2/M^2$ \\
  \hline
   ${\rm EQ}_{00}$  & 0.5000 & 0.5000 & 4.0    & -4.0    & 0.0 &  0.0   \\
   ${\rm NE}_{00}$  & 0.4909 & 0.7478 & 4.8348 & -3.2232 & 0.0 &  0.0   \\
   ${\rm EQ}_{+0}$  & 0.3444 & 0.5000 & 4.0    & -4.0    & 0.2 &  0.0   \\
   ${\rm EQ}_{+-}$  & 0.3444 & 0.3444 & 4.0    & -4.0    & 0.2 & -0.2   \\
   ${\rm NE}_{+-}$  & 0.3436 & 0.7140 & 4.8    & -3.2    & 0.2 & -0.2   \\
   ${\rm NEa}_{+-}$ & 0.3436 & 0.7140 & 7.2    & -4.8    & 0.2 & -0.2   \\
   ${\rm NEb}_{+-}$ & 0.3436 & 0.5496 & 4.8    & -3.2    & 0.2 & -0.4486 \\
  \hline \hline
 \end{tabular}
\end{table}
\end{center}

We use the \emph{horizon mass} $m$ to define the black hole mass ratio
$q$ on the initial slice,
\begin{equation}
q = m_1/m_2,
\end{equation}
where $m_{1}$ is the horizon mass of the lighter hole.  The symmetric
mass ratio is
\begin{equation}
\nu \equiv m_{1} \, m_{2} / (m_{1} + m_{2})^2.
\end{equation}

The horizon mass is derived from the apparent horizon's
\emph{irreducible mass}  using the
Christodoulou formula \cite{Christodoulou70}:
\begin{equation}
m^2 = m^2_{irr} + \frac{S^2}{4m^2_{irr}},
\end{equation}
where $m_{irr} = \sqrt{A_{AH}/16\pi}$ and $A_{AH}$ is the area of the
apparent horizon, which we locate using an adapted version of Jonathan
Thornburg's {\tt AHFinderDirect} code \cite{Thornburg:2003sf}.

The spin vector $\vec{S}$ applied to each hole in the Bowen-York data
prescription (\ref{eqn:BYspin}) is in fact the ADM angular momentum of
that hole -- the total angular momentum of the spatial slice as
measured by an ADM integral at infinity if no other sources were
present. Thus these parameters represent a global quantity.

The more standard definition of astrophysical black-hole spin,
however, is a local one. The Kerr solution is parametrized by the
black hole horizon mass $m$, and a spin parameter $a$ restricted to
the range of values $[0,m]$. Then, $a = |\vec{S}| / m$. We note,
however, that it has proved impossible to bring Bowen-York spinning
data to an $a$ value of more than $\sim 0.927 m$, significantly below
the maximal Kerr value \cite{Dain:2002ee}.

Even in the head-on case, with spin vectors orthogonal to the
separation vector, there is much freedom in the parameters describing
the initial data. In particular, we have chosen to scale all distances
so that the horizon mass of the lighter hole, $m_1$, is kept the same
relative to the grid spacing (ensuring a common level of resolution of
that hole's features). Within this restriction, we have also striven
to maintain the same proper horizon-to-horizon separation $l$ between
the holes.

\section{Methodology}
\label{sec:methodology}

These initial data sets were evolved using the moving puncture method
as implemented in the {\tt Hahndol} code \cite{Baker:2006kr}. We used
standard Baumgarte-Shapiro-Shibata-Nakamura (BSSN)
\cite{Baumgarte99,Shibata95} evolution equations, with the addition of
dissipation terms as in \cite{Huebner99} and constraint-damping terms
as in \cite{Duez:2004uh}, in order to ensure robust stability.  Our
gauge conditions are 1+log lapse slicing and a hyperbolic Gamma-driver
shift condition, as used in \cite{vanMeter:2006vi}. Here we take $\eta
= 2.0$, and the initial lapse shape chosen is $\alpha_{init} =
\psi_0^{-4}$, where $\psi_0$ is the initial conformal factor; this
gives a steeper fall-off near the punctures than the $\psi_0^{-2}$
used in the evolutions of \cite{Baker:2006kr}.  Time integration was
carried out with a fourth-order Runge-Kutta algorithm, and spatial
derivatives with fourth-order-accurate finite differencing stencils.
{We employed a second-order-accurate Sommerfeld condition on the outer
boundary at $256M$; physical modes propagating in from the outer
boundary will then take $\sim 190M$ to reach the outermost extraction
sphere at $60M$, while gauge modes propagating at $\sqrt{2}$ times the
speed of light will take $\sim 140M$ to reach this same sphere. As
will be seen, the bulk of the merger radiation and associated momentum
flux will have passed through the extraction sphere by this time.}  We
used adaptive mesh refinement (AMR) implemented via the software
package {\tt PARAMESH}
\cite{paramesh,paramesh2,paramesh3,parameshMan}, with
fifth-order-accurate interpolation between refinement regions (as
preservation of fourth-order accuracy in the bulk demands better than
fourth-order-accurate interpolation at refinement interfaces
\cite{Imbiriba:2004tp,Bruegmann:2006at,Lehner:2005vc}).

The momentum kick of the merged black hole is calculated as the
aggregated time-integral of the momentum flux, or \emph{thrust},
$dP^i/dt$ -- a surface integral of the squared time-derivative of the
radiative Weyl scalar $\psi_4$ times the unit radial vector
\cite{Newman:1981fn}:
\begin{eqnarray}
\frac{dP^i}{dt} &=& \lim_{R \rightarrow \infty} \left\{ \frac{R^2}{4 \pi} \oint d\Omega \frac{x^i}{R} \left| \int^t_{t_0} dt' \, \psi_4 \right|^2 \right\} \label{eqn:thrust_formula}\\
\Delta P^i &=& \int^t_{t_0} dt \frac{dP^i}{dt'}, \label{eqn:kick_formula}
\end{eqnarray}
where $t_0$ is the initial time in the simulation (all time integrals
above should properly be from $t' = -\infty$; our implementation
neglects thrust contributions from before $t_0$).  {In fact, for
reasons discussed in the next section, we must delay the integration
start time $t_0$ to when the thrust passes through zero. As
(\ref{eqn:thrust_formula}) is quadratic in the waveform $\psi_4$, the
kick can be expected to be weak.}  To perform the angular integration
in (\ref{eqn:thrust_formula}), we use the second-order Misner
algorithm described in \cite{Misner:1999ab,Fiske:2005fx}.


\section{Code Calibration}
\label{sec:tests}

We have established the convergence of the {\tt Hahndol} code in prior
publications \cite{Baker:2006kr,Baker:2006vn}.  To these previously
published results, we have added only one qualitatively new aspect:
spin on the pre-merger holes. Thus we restrict our discussion of
convergence here to a study of the convergence properties of a sample
spinning data set with equal masses and anti-aligned spins, ${\rm
EQ}_{+-}$.

Our main evolutions were performed at a maximum resolution of $M/32$
in the vicinity of the pre-merger holes. We emphasize here that, as we
have maintained a smaller horizon mass of $m \approx 0.5 M$ in all
runs, the grid spacing realized near the smaller hole will always be
close to $m/16$.  {This yields corresponding grid spacings of $h=M$,
$h=2M$ and $h=8M$ at the extraction surfaces $R_{\rm ext} = 30M$,
$R_{\rm ext} = 60M$ and the outer boundary, respectively.  Relative to
this standard resolution ($M/32$), we have added two higher
resolutions ($M/40$ and $M/48$).}

Fig. \ref{constraint_converge} indicates the rate of convergence of
the Hamiltonian and momentum constraints in two regions of interest:
the strong-field region surrounding the holes (left panels), and the
refinement level containing our $R_{\rm ext} = 60M$ wave-extraction
sphere (right panels).  In our simulations, the Hamiltonian constraint
exhibits greater than third-order convergence, and the momentum
constraint exhibits second-order convergence through most of the
duration of the runs.  This may be partly due to our
second-order-accurate initial data solver; however errors of
lower-than-expected order have also been attributed to ``leakage''
from the punctures \cite{Campanelli:2006fy,Brown:2007nt}.  At late
times the apparent convergence of the momentum constraint drops to a
rate between first and second order; we believe this is because of a
high-frequency gauge pulse propagating outward until it is
insufficiently resolved at the grid resolution achieved in the outer
zones of our coarser run.  However, this gauge pulse does not
significantly affect the thrust computation, as evidenced by the
invariance with extraction radii demonstrated in
Sec.~\ref{sec:results}.

In Fig. \ref{psi22_converge}, we illustrate the convergence of the
dominant $(2,2)$ mode of the Weyl scalar ${\rm Re}(\psi_{4})$,
extracted at $R_{\rm ext} = 60M$.  For comparisons, we have lined up
the main amplitude peak in time across resolutions, and rotated the
waveform phase to zero at this point (similar to the \emph{maquillage}
used in inspiral waveform comparisons \cite{Buonanno:2006ui}).
Resolution-dependent errors are higher during the unphysical
high-frequency Bowen-York pulse (see next section); by $\sim 100M$,
when the physical merger radiation reaches the detector, errors are
consistent in overall magnitude with second-order convergence, and
compare favorably with the equivalent plots in Fig. 3 of
\cite{Baker:2006yw}.

\begin{widetext}
\begin{center}
\begin{table}[t]
 \caption{{Derived initial parameters in terms of the fiducial mass
 $M$.}  $m_1$ is the horizon mass of the lighter hole.  $q$ is the
 mass ratio defined by horizon masses.  $a_A = |S_A|/m_A$ is the
 approximate Kerr spin parameter of hole $A$. $l$ is the proper
 horizon-horizon separation of the holes.  $M_{ADM}$ was calculated at
 a finite coordinate distance from the origin ($60 M$ for all but run
 ${\rm EQ}_{+0}$){, using Eq. (12) of \cite{Omurchadha74}. Errors in
 horizon masses $m_1$ and $m_2$ are $\sim 1.5\%$ (see text), which
 propagate into derived quantities $q$, $\nu$, $a_1$, $a_2$,
 $a_1/m_1$, and $a_2/m_2$. ADM mass errors enter at the fourth
 significant digit.}}
\label{table:idderived}
 \begin{tabular}{c| c c c c c c c c c c}
  \hline  \hline
   Run      & $m_1/M$ & $m_2/M$ & $q$ & $\nu$ & $a_1$ & $a_2$ & $a_1/m_1$ & $a_2/m_2$ &  $l/M$ & $M_{ADM}/M$ \\
  \hline
   ${\rm EQ}_{00}$  & $0.514$  & $0.514$  & 1.0               & 0.250             & 0.0               & 0.0   & 0.0   & 0.0   & 12.24   & 1.0   \\
   ${\rm NE}_{00}$  & $0.514$  & $0.771$  & 0.667 & 0.240 & 0.0               & 0.0   & 0.0   & 0.0   & 12.24   & 1.24 \\
   ${\rm EQ}_{+0}$  & $0.516$  & $0.514$  & 1.004   & 0.250 & 0.389 & 0.0   & 0.758 & 0.0   & 12.24   & 1.0   \\
   ${\rm EQ}_{+-}$  & $0.514$  & $0.514$  & 1.0   & 0.250 & 0.389 & 0.389 & 0.758 & 0.758 & 12.4 $\pm$ 0.2  & 1.0   \\
   ${\rm NE}_{+-}$  & $0.516$  & $0.773$  & 0.668 & 0.240 & 0.388 & 0.259 & 0.752 & 0.335 & 12.6 $\pm$ 0.2  & 1.24 \\
   ${\rm NEa}_{+-}$ & $0.513$  & $0.764$  & 0.672 & 0.240 & 0.390 & 0.262 & 0.759 & 0.342 & 17.0 $\pm$ 0.2  & 1.25 \\
   ${\rm NEb}_{+-}$ & $0.516$  & $0.784$  & 0.658 & 0.239 & 0.388 & 0.572 & 0.752 & 0.730 & 13.0 $\pm$ 0.2  & 1.26 \\
  \hline \hline
 \end{tabular}
\end{table}
\end{center}

\begin{center}
\begin{figure}[ht]
\includegraphics*[scale=.6]{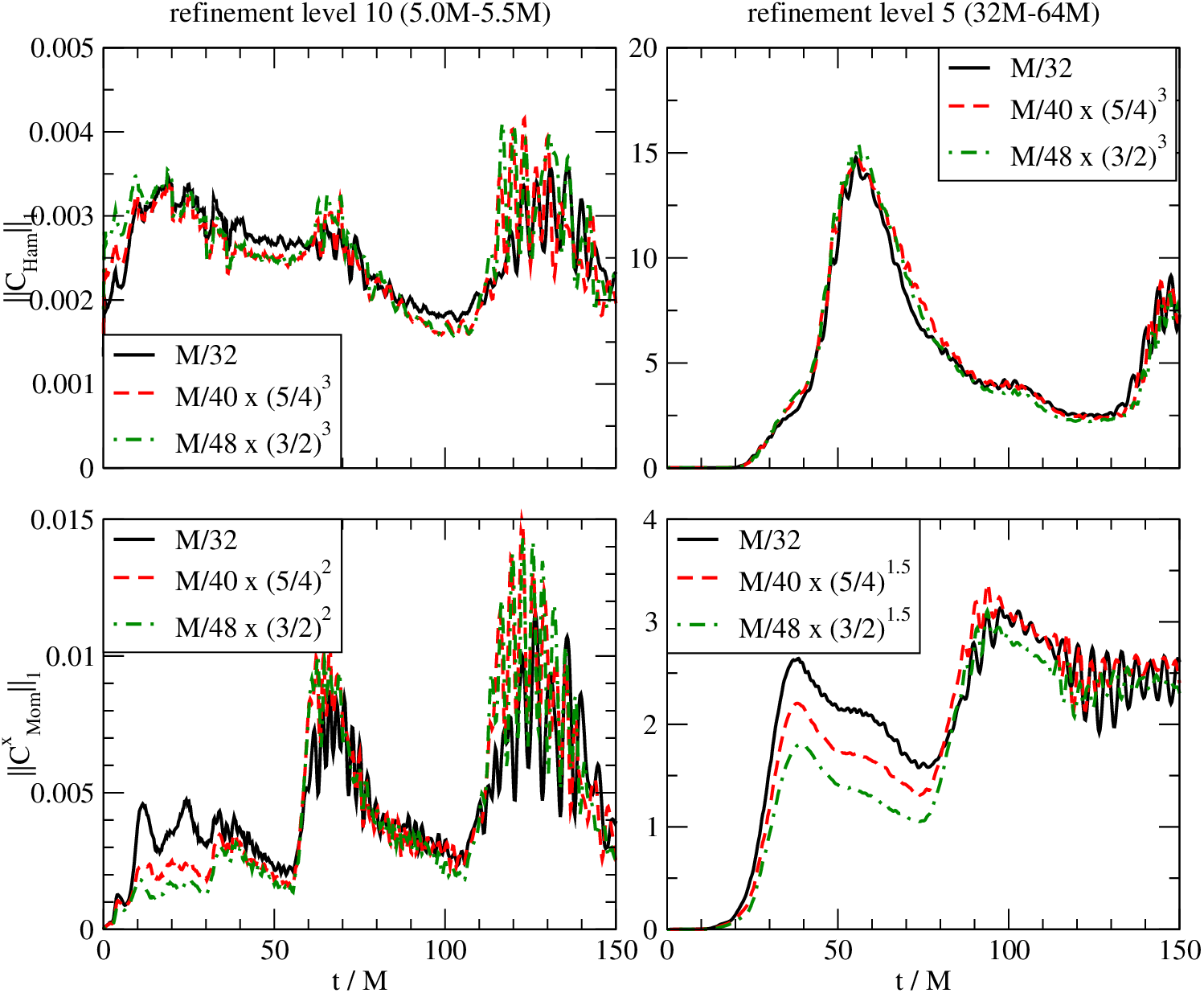}
\caption{L1-norm convergence of Hamiltonian (top) and momentum ($x$ component; bottom)
constraints for the ${\rm EQ}_{+-}$ run, in two regions of the simulation grid: just outside the
horizons (left),and in the primary radiation extraction region
(right). The Hamiltonian constraint displays third-order convergence in
both regions, while the momentum constraint convergence drops to 1.5-order
in the extraction region.}
\label{constraint_converge}
\end{figure}
\end{center}
\end{widetext}

\begin{center}
\begin{figure}[ht]
\includegraphics*[scale=.35]{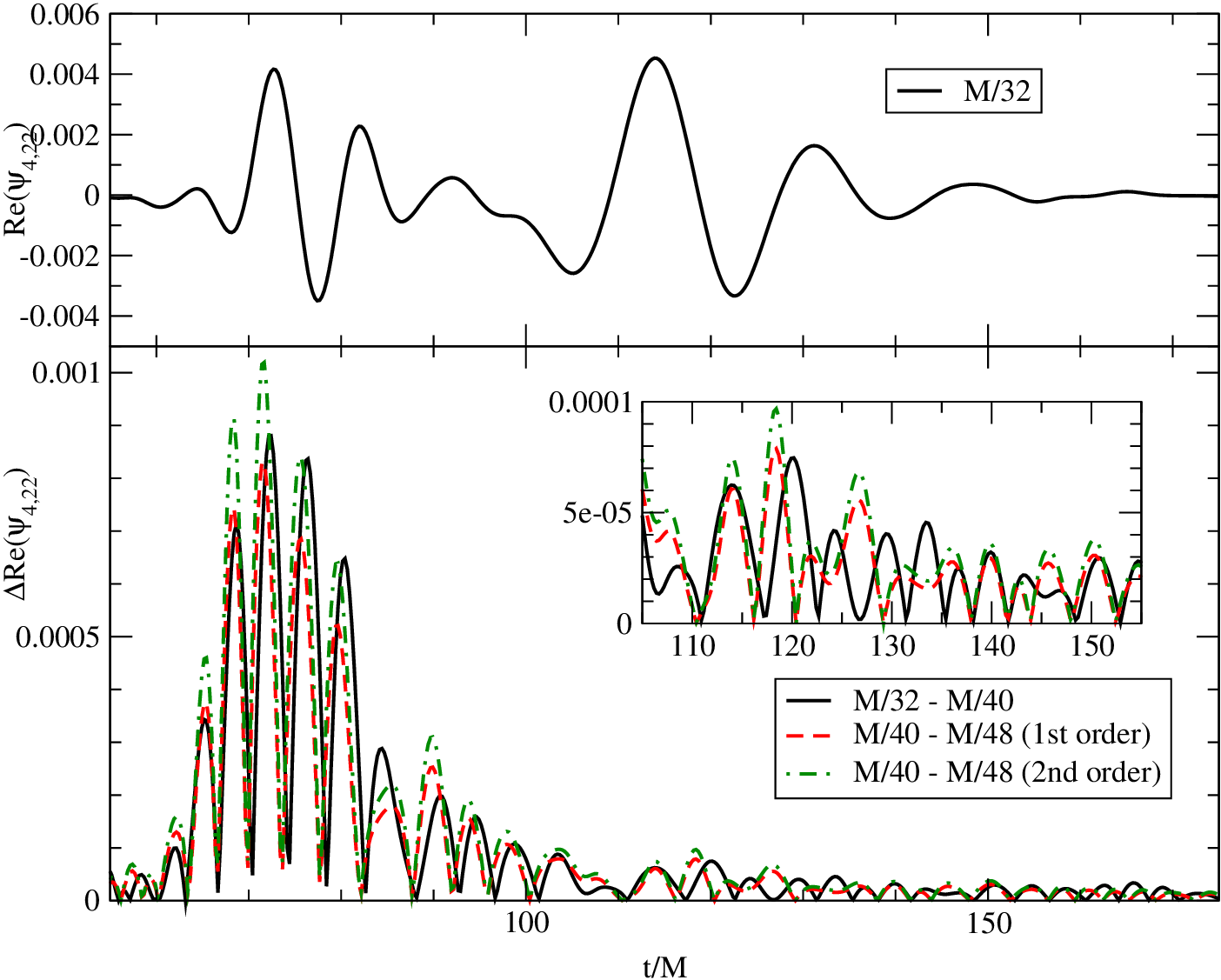}
\caption{Convergence of ${\rm Re}(\psi_{4})$, the real part of the dominant $(2,2)$
mode of the waveform for the ${\rm EQ}_{+-}$ run. The upper panel
shows the thrust for central resolution $M/32$, extracted at $R_{\rm
ext} = 60M$; the lower panel shows the absolute differences $M/32 -
M/40$ (solid line) and $M/40 - M/48$, the latter scaled to compare
with coarse-medium for 1st- and 2nd-order convergence. The inset
focuses on the physical part of the waveform, which shows much smaller
resolution errors.}
\label{psi22_converge}
\end{figure}
\end{center}

\begin{center}
\begin{figure}[ht]
\includegraphics*[scale=.35]{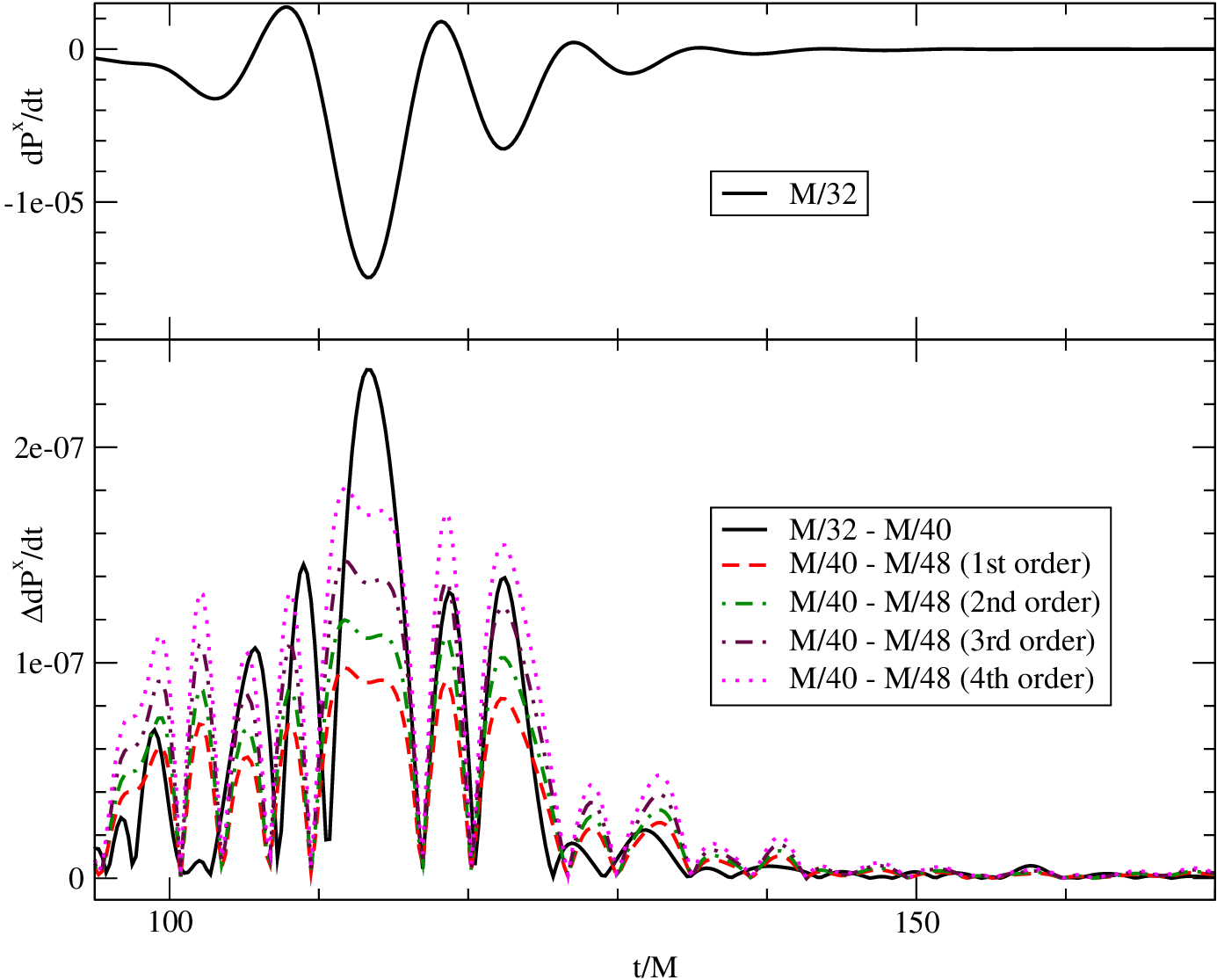}
\caption{Convergence of transverse momentum thrust $dP^x/dt$ for the ${\rm EQ}_{+-}$
run. The upper panel shows the thrust for central resolution $M/32$,
extracted at $R_{\rm ext} = 60M$; the lower panel shows the absolute
differences $M/32 - M/40$ (solid line) and $M/40 - M/48$, the latter
scaled to compare with coarse-medium for 1st- and 2nd-order
convergence.}
\label{kick-dot_converge}
\end{figure}
\end{center}

In Fig. \ref{kick-dot_converge}, we illustrate the convergence of the
momentum thrust for our data set ${\rm EQ}_{+-}$, extracted at $R_{\rm
ext} = 60M$ (as for Fig. \ref{psi22_converge}), beginning at the
integration start time $t_0 = 90.5M$.  To compare the different data
sets, we re-zero and rescale the time axis to overlay the two largest
physical peaks for all three resolutions.  The differences have again
been scaled for first- and second-order convergence. The thrust
displays better than second-order convergence for most of the physical
thrust, but has a significant excursion around $t=102M$. Nevertheless,
the relative errors in the thrust are $\sim 1\%$ (compare with upper
panel).

\subsection{Bowen-York Radiation Pulses}
\label{ssec:BY_pulse}

Bowen-York black hole initial data contains a non-negligible amount of
``false radiation,'' even after the Hamiltonian constraint has been
solved. This radiation is unphysical in the sense that it is a
reflection of the unphysical approximations, such as conformal
flatness, made in generating the initial data (\emph{c.f.} 
\cite{Hannam:2006zt}).

To estimate the approximate duration of this unphysical radiation
pulse, we assume that each hole relaxes to a Kerr hole (of
approximately the same mass and spin) through the emission of
quasinormal modes (QNMs). The most slowly damped of these will have an
``e-folding'' time $\tau_e \approx 12 m$ for a hole of mass $m$
\cite{Berti:2005ys}. Then a pulse will have decayed by two orders of
magnitude after $\sim 55 m$ -- that is, $\sim 30 M$ for our equal-mass
(${\rm EQ}$) data sets, and $\sim 45 M$ for our unequal-mass (${\rm
NE}$) data.

This Bowen-York ``pulse'' is concentrated close to each hole's
horizon, and much of it falls into its parent hole{, resulting in an
increase in apparent-horizon mass of $\sim 1\%$}; however, what
radiation escapes to our detection spheres will mix with the physical
radiation we are generally interested in. Nevertheless, since the
pulse is released immediately, and is of short duration, it should be
easily identifiable in the total waveform.

\begin{center}
\begin{figure}[ht]
\includegraphics*[scale=.35]{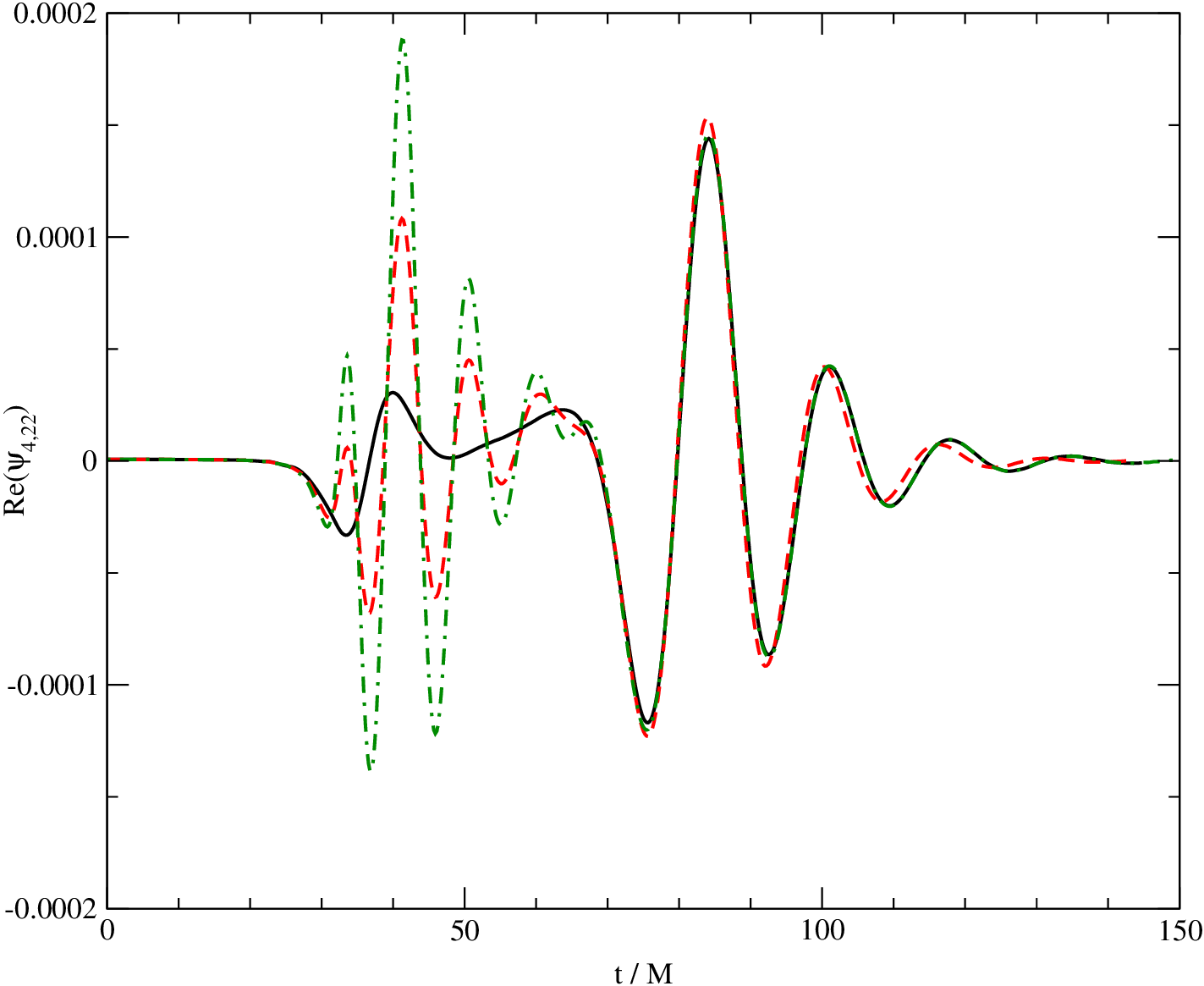}
\caption{Dominant $(l=2,m=2)$ mode of the waveform $\psi_4$ for the
  three equal-mass configurations: ${\rm EQ}_{00}$ (solid/black),
  ${\rm EQ}_{+0}$ (dashed/red), and ${\rm EQ}_{+-}$
  (dot-dashed/green).  These data were extracted at coordinate
  distance $R_{\rm ext} = 30 \, M$.}
\label{EQ_WF}
\end{figure}
\end{center}

To illustrate, we show in Fig.~\ref{EQ_WF} the dominant $(l=2,m=2)$
spin-2-weighted spherical harmonic mode of the ``outgoing'' Weyl
scalar $\psi_4$ (extracted at $R_{\rm ext} = 30 M$), for three
different equal-mass configurations: zero-spin (${\rm EQ}_{00}$),
single-spin (${\rm EQ}_{+0}$) and double-spin (${\rm
EQ}_{+-}$). Despite the differences in spins, these three waveforms
agree well in the later ``physical'' part of the signal, with stronger
differences occurring in the nonphysical BY pulse. In the zero-spin
case, the BY pulse is {small}, but it is significant for ${\rm
EQ}_{+0}$ and ${\rm EQ}_{+-}$.

This same pulse will yield a non-zero contribution to the momentum
kick. Again, this should be easy to identify, as it will result in a
plateau early in the run. Having identified a time $t_0$ at which the
BY pulse and its ``pseudo-kick'' are finished, we remove this effect
by beginning the integration of (\ref{eqn:thrust_formula}) from that
point.


\section{Results}
\label{sec:results}

We group our simulations into three categories. First we consider the
effect of unequal masses alone, where the kick experienced should be
along the line connecting the holes. We then add spins to holes of
equal horizon mass, to explore the dependence of kicks on pure spin
anisotropy; this configuration should yield kicks orthogonal to the
line connecting the holes. Finally we consider the combination of both
spin and unequal masses, to explore the interrelation of the two
mechanisms generating kicks.

\subsection{Unequal masses without spin}

The head-on collision of two holes of unequal masses has been
considered several times in the past, both with close-limit analysis
(CLA) \cite{Andrade:1997pc} and with fully numerical evolutions in 2D
\cite{Anninos98a}. The CLA results (performed with Misner two-sheeted
data) provide a definite prediction of preferential radiation of
linear momentum; the numerical results seemed to confirm this, and
indicated where the CLA fails. The numerical results were for proper
separations far less than ours; extrapolating their results to the
larger separations we treat here indicates that a longitudinal kick
$\sim 10 {\rm km/s}$ would be produced.

\begin{center}
\begin{figure}[ht]
\includegraphics*[scale=.40]{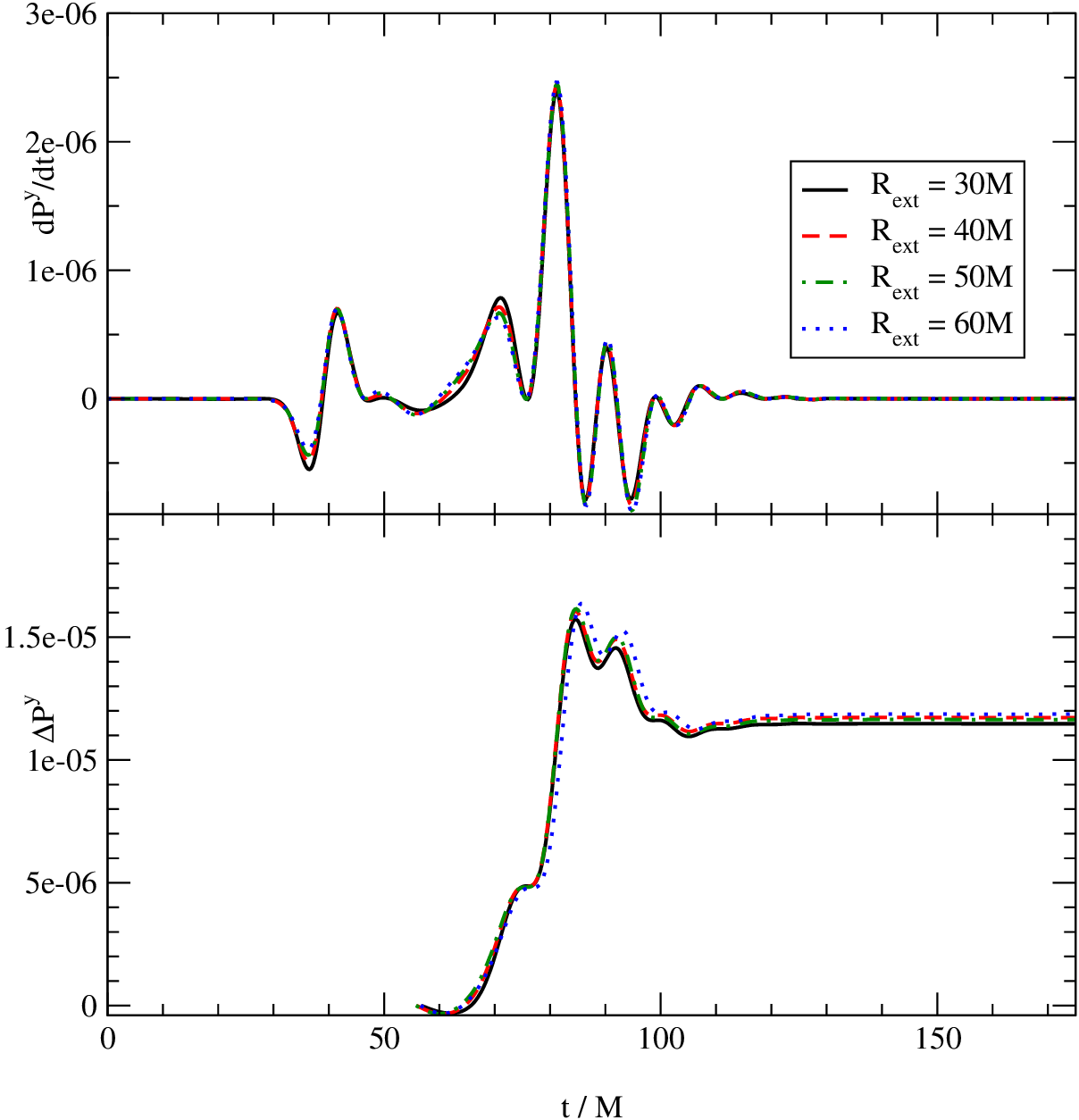}
\caption{Longitudinal thrust $dP^y/dt$ (top) and kick $\Delta P^y$
(bottom) for ${\rm NE}_{00}$, at extraction radii $R_{\rm ext}$ of $30
M$, $40 M$, $50 M$ and $60 M$, where the latter three have been
time-shifted by $10.6 M$, $21.0 M$ and {$31.4 M$,
respectively, to
align the highest thrust peak with the $30M$ case (consistent with
time-shift formula Eq. (14) in \cite{Fiske:2005fx}).}}
\label{NE00_Py_dot_int}
\end{figure}
\end{center}

In Fig. \ref{NE00_Py_dot_int}, we plot the $y$-component of the thrust
$d\vec{P}/dt$ from Eq. (\ref{eqn:thrust_formula}), and its time
integral from the ${\rm NE}_{00}$ run. Note that the $x$ and $z$
components of $d\vec{P}/dt$ are zero by symmetry: the kick is
longitudinal, i.e. along the axis of collision ($y$ axis). The final
kick velocity, given in Table \ref{table:results}, is $\sim 2.7 {\rm
km/s}$.
The final kick obtained from Fig. \ref{NE00_Py_dot_int} is highly
consistent across extraction radii, with a spread between the $R_{\rm
ext} = 30 M$ and $R_{\rm ext} = 40 M$ values of $\sim 0.5 \%$. We note
however that the difference between $R_{\rm ext} = 50 M$ and $R_{\rm
ext} = 60 M$ is larger than that between $R_{\rm ext} = 40 M$ and
$R_{\rm ext} = 50 M$. This appears to be because the extraction
surface at $R_{\rm ext} = 60 M$ lies in a region of less grid
refinement than the three closer surfaces.

\subsection{Equal masses with spin}

We now turn to simulations of spinning black holes.  We ran two
configurations (${\rm EQ}_{+0}$,${\rm EQ}_{+-}$), as indicated in
Table \ref{table:idparams}. As only one particle is spinning in the
former case, we should see only ``spin-orbit'' effects between that
spin and the motion of the particle. In the latter case, with two
spins, we might also see ``spin-spin'' effects; however PN predictions
for the latter are of higher PN order than for spin-orbit
\cite{Kidder95a}\footnote{ 
For the ${\rm EQ}_{+0}$ run presented here, we have used slightly
different choices in our numerical evolution: advection terms employed
a third-order upwinding scheme, guardcell-filling was
fourth-order-accurate, time stepping was performed with a second-order
Crank-Nicholson method, no dissipation was used, and the initial lapse
function shape was $\alpha_{init} = \psi_{BL}^{-2}$, as in the
evolutions of \cite{Baker:2006kr}. For this data alone, the physical
outer boundary was at $128 M$, but the outermost extraction here was at
$R_{\rm ext} = 30 M$.  {From direct comparison of the thrusts obtained
from the two evolution schemes for the ${\rm EQ}_{+-}$ data, we have
determined that the only measurable difference is a gauge-driven $1M$
delay in the arrival of wavefronts at the extraction sphere.}}

\begin{center}
\begin{figure}[ht]
\includegraphics*[scale=.40]{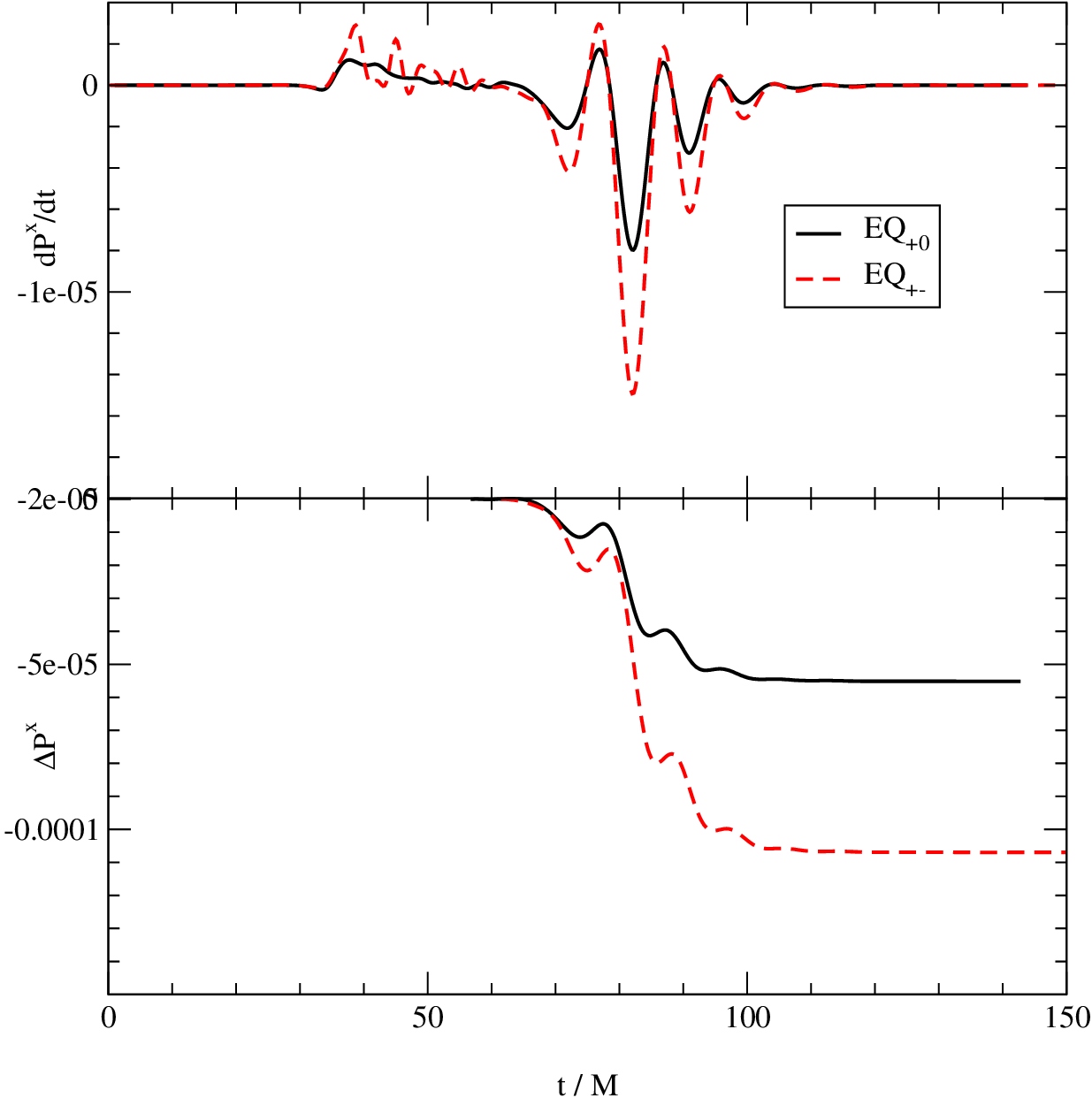}
\caption{Transverse thrust $dP^x/dt$ and kick $\Delta P^x$ for the spinning,
equal-mass cases ${\rm EQ}_{+0}$ (black/solid) and ${\rm EQ}_{+-}$
(red/dashed), at extraction radii $R_{\rm ext}$ of $30 M$, where the
${\rm EQ}_{+-}$ data have been time-shifted by $1M$ to compensate for
gauge differences.}
\label{EQ_Px_dot_int}
\end{figure}
\end{center}

In Fig. \ref{EQ_Px_dot_int} we show the transverse kick $\Delta P^x$
observed for both equal-mass data sets. We note that the magnitude of
both the BY pulse and the final kick for the ${\rm EQ}_{+-}$ data are
roughly double those for ${\rm EQ}_{+0}$ (see Table
\ref{table:results}). This indicates that the total kick is close to a
simple sum of the individual kicks. Thus any spin-spin contribution to
the total kick is negligible in comparison to the spin-orbit term.

\subsection{Unequal masses with spin}

Having considered collisions between non-equal mass and non-spinning
black holes and equal mass and spinning black holes, we combine both
spin and mass ratio in a few cases, the ${\rm NE}_{+-}$, ${\rm
NEa}_{+-}$, and ${\rm NEb}_{+-}$ runs.

\begin{center}
\begin{figure}[ht]
\includegraphics*[scale=.40]{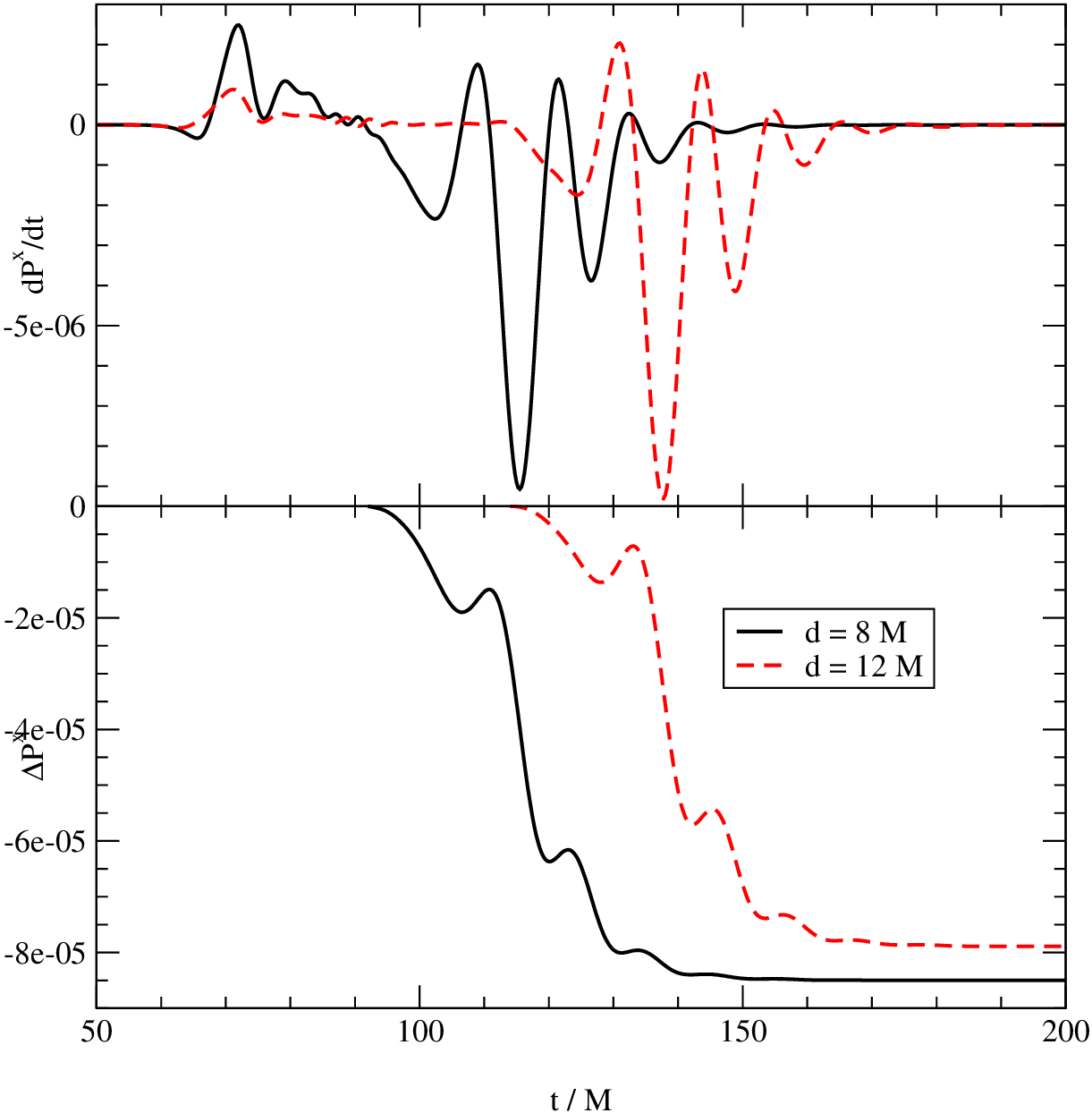}
\caption{Transverse thrust $dP^x/dt$ (top) and kick $\Delta P^x$ 
(bottom) for unequal-mass-with-spin data, at coordinate separations $8
M$ (${\rm NE}_{+-}$; black/solid) and $12 M$ (${\rm NEa}_{+-}$;
red/dashed). These data were extracted at $R_{\rm ext} = 60 M$.}
\label{large_sep_Pxint}
\end{figure}
\end{center}

We show in the upper panel of Fig. \ref{large_sep_Pxint} the thrust
for the ${\rm NE}_{+-}$ case. We note that unlike the simpler cases
before, the BY pulse and physical signal are \emph{not} obviously
separated in the thrust. As a result, we may expect to have trouble
determining where to begin the kick integrations.

In principle, a larger initial separation would mean more time between
the BY pulse -- emitted at $t = 0$ and lasting $\sim 43 M$ (see
discussion in Section \ref{ssec:BY_pulse}) -- and the bulk of the
physical radiation, which only becomes large near merger. The main
physical kick will certainly be different in magnitude between the two
cases, and even the BY pulse magnitude may differ, depending on the
dependence of the spectrum on the initial separation of the
binary. Nevertheless, we \emph{can} expect the BY pulse
\emph{duration} to be the same, as this is determined by the
leading quasinormal frequencies of each hole, and these frequencies
are fully determined by the mass and spin of each hole, which is a
constant between the two cases. In Fig. \ref{large_sep_Pxint}, we show
$dP^x/dt$ and $\Delta P^x$ for ${\rm NE}_{+-}$, with a coordinate
separation of $8 M$, and ${\rm NEa}_{+-}$, with a coordinate
separation of $12 M$.

The extra initial separation {(${\rm NEa}_{+-}$)} leads to a very
similar physical thrust, delayed $\sim 22 M$ relative to the original
separation {(${\rm NE}_{+-}$)}. This physical thrust begins at $\sim
114 M$ for the detector at $R_{\rm ext} = 60 M$. Integrating forward
from any time up to $\sim 20 M$ before this will result in the same
transverse kick for the ${\rm NEa}_{+-}$ case. To ensure that as much
of the physical kick as possible is obtained in the smaller-separation
case, we integrate the ${\rm NE}_{+-}$ thrust from $t_0 = 114 M - 22 M
= 92 M$ at $R_{\rm ext} = 60 M$.

\begin{center}
\begin{figure}[ht]
\includegraphics*[scale=.40]{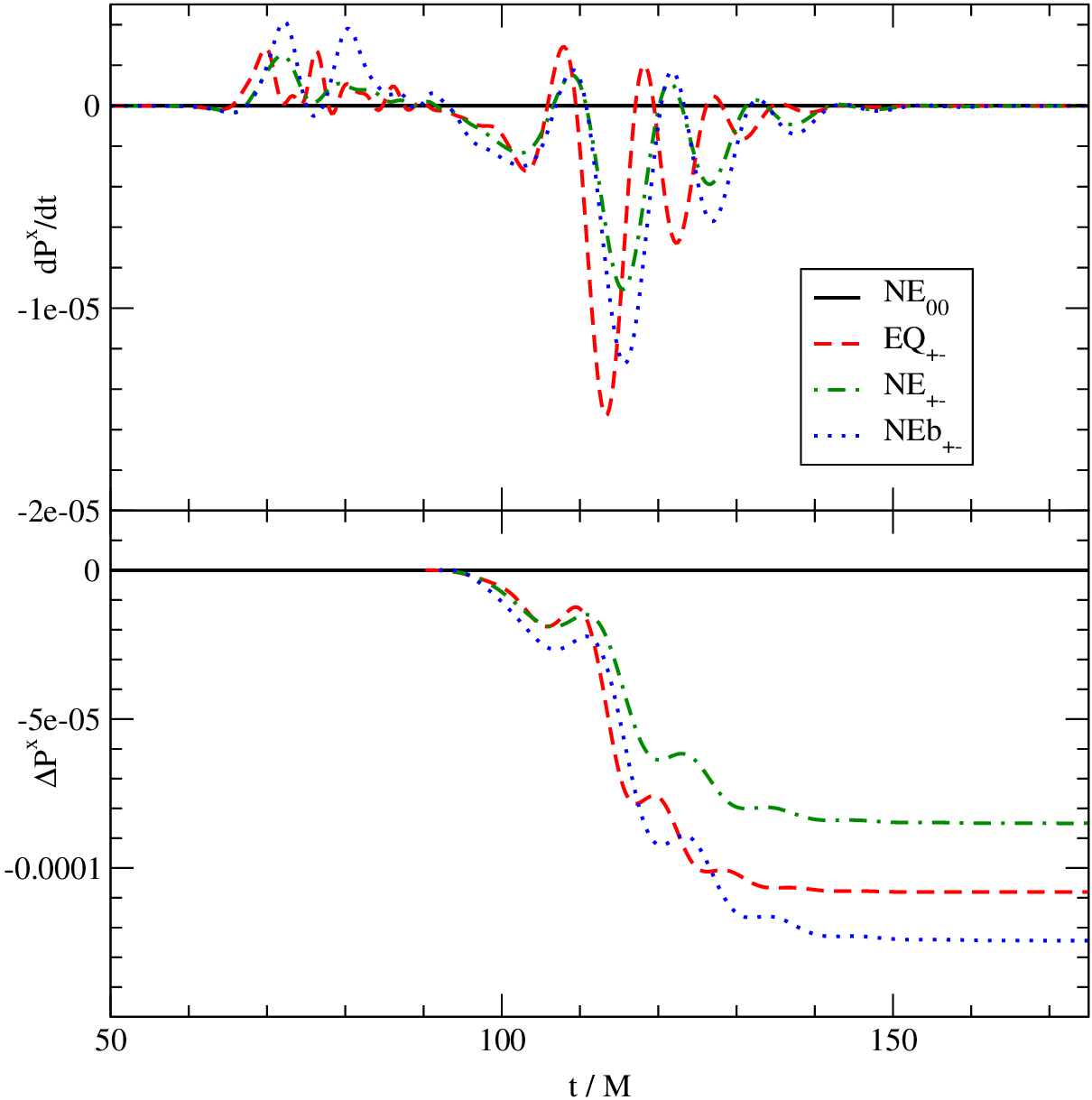}
\caption{Transverse thrust $dP^x/dt$ (top) and kick $\Delta P^x$
  (bottom) for ${\rm NE}_{00}$ (black), ${\rm EQ}_{+-}$ (red), and
  ${\rm NE}_{+-}$ (green), and ${\rm NEb}_{+-}$ (blue). These data
  were extracted at $R_{\rm ext} = 60\,M$.}
\label{final_Px_dot_int}
\end{figure}
\end{center}

\begin{center}
\begin{figure}[ht]
\includegraphics*[scale=.40]{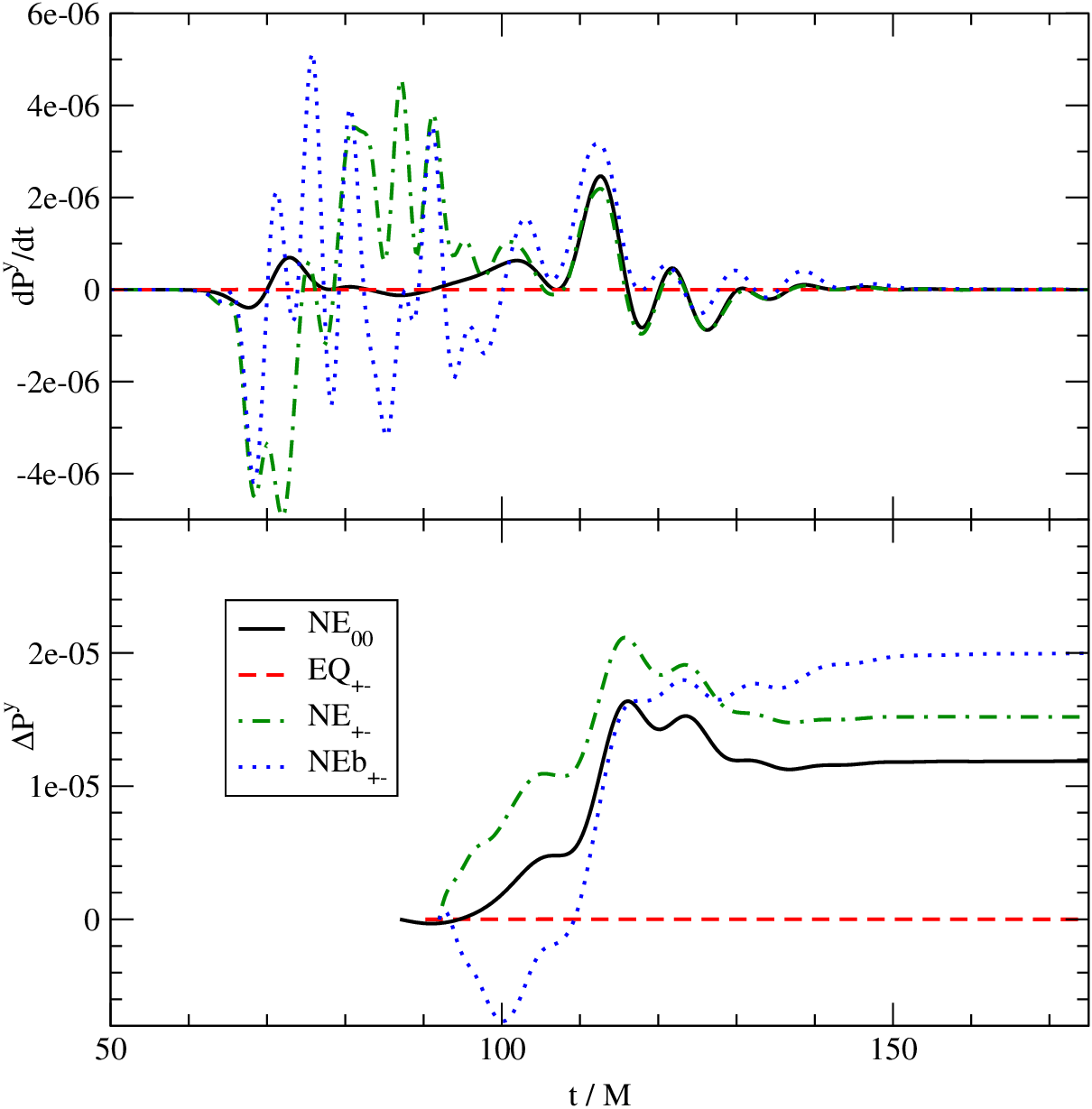}
\caption{Longitudinal thrust $dP^y/dt$ (top) and kick $\Delta P^y$ 
  (bottom) for ${\rm NE}_{00}$ (black), ${\rm EQ}_{+-}$ (red), and
  ${\rm NE}_{+-}$ (green), and ${\rm NEb}_{+-}$ (blue). These data
  were extracted at $R_{\rm ext} = 60 \, M$.}
\label{final_Py_dot_int}
\end{figure}
\end{center}

In Figs. \ref{final_Px_dot_int} and \ref{final_Py_dot_int} we present
the momentum kicks in the transverse ($x$) and longitudinal ($y$)
directions for the ${\rm NE}_{+-}$ data set, comparing with the kicks
seen from the ${\rm NE}_{00}$ and ${\rm EQ}_{+-}$ cases before.

Taking the transverse direction first, we see from
Fig. \ref{final_Px_dot_int} that the momentum kick from ${\rm
NE}_{+-}$ is significantly \emph{smaller} than that of ${\rm
EQ}_{+-}$. The spin angular momentum $S$ present in each case is the
same, indicating that the kick is not a function of $S$. In contrast,
the momentum kick observed from ${\rm NEb}_{+-}$ is much larger.

The physical momentum kicks from these runs are presented in Table
\ref{table:results}. We note that the ratio of transverse momentum kicks
$\Delta P^x$ between the cases ${\rm NE}_{+-}$ and ${\rm NEb}_{+-}$ is
roughly $2/3$, consistent with the PN-derived ratio {(see
Eq. (\ref{eqn:PN_PSO_simp}) below)}.

In addition to finite differencing inaccuracy, for these cases we find
that the high slope in the thrust for these cases yields a
non-negligible error associated with the choice in integration
starting point $t_0$.  {As before, we assess a timing-related error by
integrating the thrust from $t_0$ and from $t_0 \pm 5M$. The resulting
central kicks and errors are given in Table \ref{table:results}.}

With this in mind, we turn to the longitudinal direction. Here we see
that the initial BY burst is comparable in magnitude to the physical
later part. Using for consistency the same integration starting time
as in the transverse direction, we present the longitudinal kicks in
Table \ref{table:results}. {It is clear that the $t_0$ uncertainty
leads to large relative errors for the longitudinal kicks.}  To this
uncertainty, all longitudinal kicks are consistent with the zero-spin
case, ${\rm NE}_{00}$.

\subsection{Summary of Results}

We draw together in Table \ref{table:results} the kicks observed in
each of our simulations.  We can develop some expectations for the
comparative results by referencing post-Newtonian theory estimates of
the radiative linear momentum loss due to unequal masses and
spin-orbit coupling. For example, adapting Eqs. (3.31a-b) of
\cite{Kidder95a} to radial infall along the $y$ axis, we find:
\begin{eqnarray}
\dot{\vec{P}}_N &=& \frac{16}{105} \frac{\delta m}{m_T} \nu^2 \left( \frac{m_T}{r} \right)^4 \dot{r} \left( - \dot{r}^2 + \frac{2 m_T}{r} \right) \hat{e}_y \label{eqn:PN_PN_simp}, \\
\dot{\vec{P}}_{SO} &=& - \frac{16}{15} \frac{\nu^2 \dot{r}^2}r \left( \frac{m_T}{r} \right)^4 ( a_1 + a_2 ) \hat{e}_x, \label{eqn:PN_PSO_simp}
\end{eqnarray}
where $m_T = m_1 + m_2$, $\delta m = m_1 - m_2$, and $r$ is the
spatial separation of the two particles, with $\dot{r} < 0$.

In studies of spin such as the one carried out here, whether physical
effects scale with angular momentum $S$, the Kerr parameter $a \equiv
S/m$ or the dimensionless number $a/m$ is an open question. The
post-Newtonian result of Eq.~(\ref{eqn:PN_PSO_simp}) predicts a thrust
(and hence kick) that scales with $a$ for each hole. We have tried to
address this uncertainty in our choice of data sets.

\begin{center}
\begin{figure}[ht]
\includegraphics*[scale=.35]{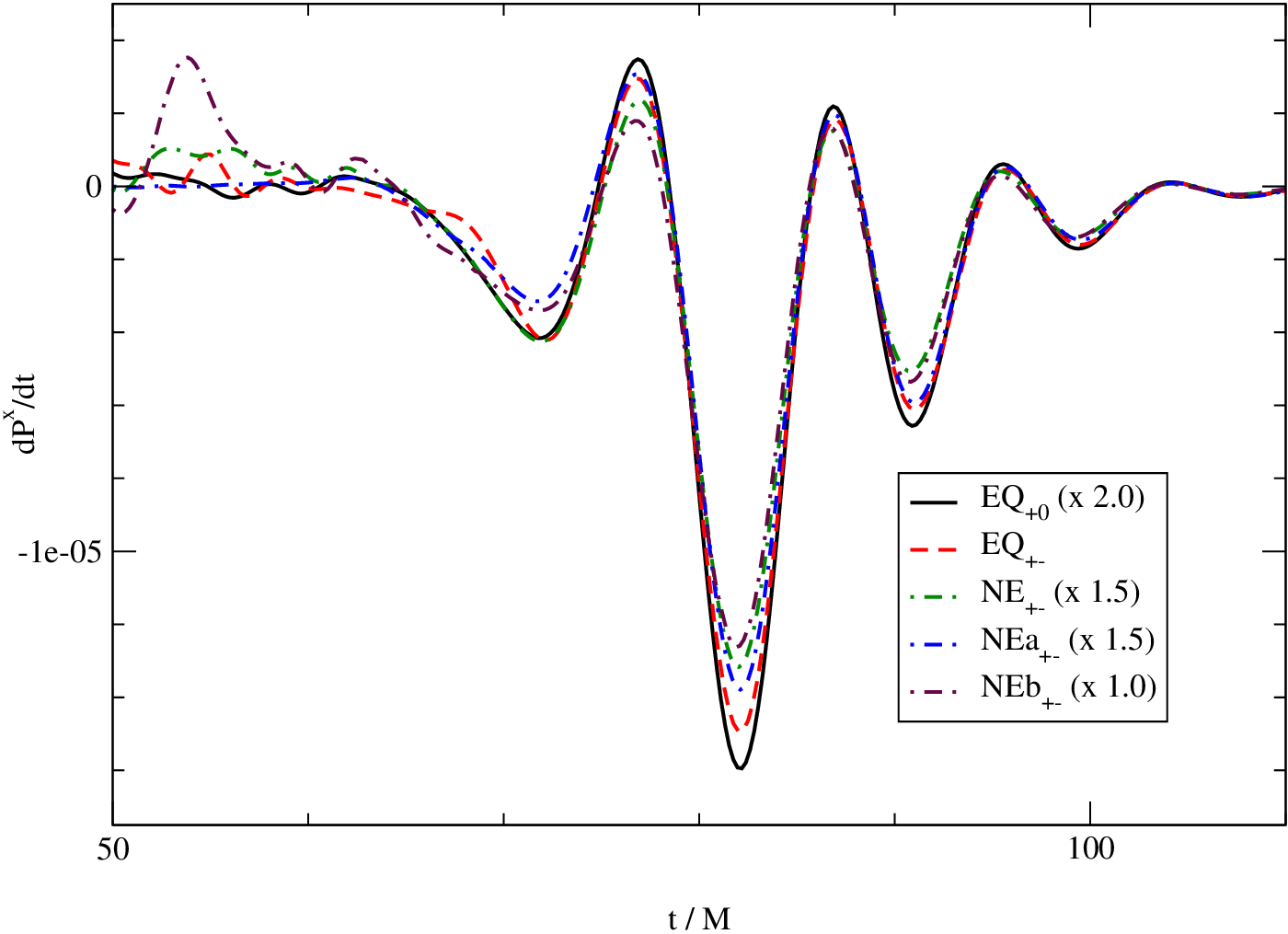}
\caption{Transverse thrust $dP^x/dt$ for all spinning data sets:
  ${\rm EQ}_{+0}$, ${\rm EQ}_{+-}$, ${\rm NE}_{+-}$, ${\rm NEa}_{+-}$,
  and ${\rm NEb}_{+-}$. Time has been rescaled by the ADM mass for
  each data set, and translated relative to ${\rm EQ}_{+-}$ to line up
  peaks. Amplitudes have been scaled relative to ${\rm EQ}_{+-}$
  according to PN predictions. These data were extracted at $R_{\rm
  ext} = 30 \, M$.}
\label{thrust_Kidder_comp}
\end{figure}
\end{center}

It is informative to compare the transverse kick results from each of
our spinning data sets with each other, in light of post-Newtonian
predictions. Looking at Eq. (\ref{eqn:PN_PSO_simp}), we assume that
the right-hand side is more or less insensitive to the details of the rate of
infall, the important length scale then being the total mass $m_T$ of
the binary. With this in mind, the remaining freedom lies in the spin
scaling $(a_1 + a_2)/m_T$. Relative to the ${\rm EQ}_{+-}$ case, this
scaling factor is $\{ 1/2, 2/3, 2/3, 1 \}$ for cases $\{ {\rm
EQ}_{+0}, {\rm NE}_{+-}, {\rm NEa}_{+-}, {\rm NEb}_{+-} \}$.  In
Fig. \ref{thrust_Kidder_comp} we combine the post-BY-pulse part of the
transverse thrust $dP^x/dt$ of each case, with times rescaled
according to the ADM mass from Table
\ref{table:idparams}, and translated so that the peaks coincide. We have
also rescaled the thrust amplitude by the factors appropriate to the PN above
Surprisingly, after this rescaling, the
five thrusts fit very well, with a deviation of overall amplitude of
$\sim 20\%$. This appears to indicate that PN predictions of recoil
kicks have validity further into the merger regime than would be
expected.  

We have looked at the puncture tracks of the holes, which give a
coordinate speed estimate for the final hole. These agree with the
radiation-generated kick in order-of-magnitude and relative scaling
(transverse vs longitudinal velocity components), but not in detail;
in particular, we do not see a velocity plateau; given the small
recoil velocities (compared to inspiral) in question here, the
puncture-derived position may just not be accurate enough. It should
be borne in mind also that the puncture track is highly
coordinate-dependent, and has no guaranteed relationship to physical
quantities such as the kick.

We also present in Table \ref{table:final_params} estimates for the
final Kerr parameters $M_f$ and $a_f$ of the post-merger hole for each
simulation. We calculate these estimates in two different ways: First,
we compare the initial ADM energy and angular momentum of the system
with the radiated energy and angular momentum. In this we are limited
by the accuracy of the ADM energy, reliable to three significant
digits. Second, we use the $(2,2)$ mode of the radiation after the
final amplitude peak to extract a quasi-normal-mode frequency and
damping time, and hence to determine the underlying Kerr parameters,
assuming the perturbative limit. This latter method is sensitive to
the determination of an appropriate starting time for the QNM fit, as
well as the neglect of QNM overtones. The determination $a/M = 0.0252$
for the ${\rm NE}_{00}$ data set, which can contain \emph{no} angular
momentum, indicates the level of error of this method.

\begin{widetext}
\begin{center}
\begin{table}[t]
  \caption{Final integrated momentum kicks and corresponding kick
  velocities. In each case, we have removed ``BY pulse'' effects
  through deferring integration until after the passage of the
  pulse. {Quoted errors at each extraction radius are obtained by
  varying the integration starting time by $\pm 5M$.}}
\label{table:results}
 \begin{tabular}{r| r| r r| r r r}
  \hline \hline
   Run      & $t_0$ & $\Delta P^x$ & $\Delta P^y$ &   $v_x$  &   $v_y$  & $v$\\
            & $(M)$ & $(10^{-5})$  & $(10^{-5})$  & (km/s) & (km/s) & (km/s)\\
  \hline
   ${\rm NE}_{00}$  $30 M$ &  51.0 & 0.0               & $\;\;1.12 \pm 0.02$ &   0.0             & $\;\;2.71 \pm 0.05$ &  $\;\;2.71 \pm 0.05$ \\
                    $40 M$ &  61.7 & 0.0               & $1.14 \pm 0.04$ &   0.0             & $2.76 \pm 0.10$ &  $2.76 \pm 0.10$ \\
                    $60 M$ &  82.9 & 0.0               & $1.16 \pm 0.04$ &   0.0             & $2.81 \pm 0.10$ &  $2.81 \pm 0.10$ \\ \hline
   ${\rm EQ}_{+0}$  $20 M$ &  46.1 & $ -5.13 \pm 0.04$ & 0.0             & $-15.38 \pm 0.12$ & 0.0             & $\;\;15.38 \pm 0.12$ \\ 
                    $30 M$ &  56.6 & $ -5.51 \pm 0.01$ & 0.0             & $-16.52 \pm 0.03$ & 0.0             & $16.52 \pm 0.03$ \\ \hline
   ${\rm EQ}_{+-}$  $30 M$ &  59.6 & $-10.70 \pm 0.08$ & 0.0             & $-32.08 \pm 0.24$ & 0.0             & $32.08 \pm 0.24$ \\
                    $40 M$ &  70.0 & $-10.53 \pm 0.09$ & 0.0             & $-31.57 \pm 0.27$ & 0.0             & $31.57 \pm 0.27$ \\
                    $60 M$ &  90.5 & $-10.81 \pm 0.15$ & 0.0             & $-32.41 \pm 0.45$ & 0.0             & $32.41 \pm 0.45$ \\ \hline
   ${\rm NE}_{+-}$  $40 M$ &  72.0 & $ -8.47 \pm 0.31$ & $1.36 \pm 1.10$ & $-20.41 \pm 0.75$ & $3.28 \pm 2.65$ & $20.67 \pm 0.85$ \\ 
                    $60 M$ &  92.0 & $ -8.50 \pm 0.27$ & $1.52 \pm 1.36$ & $-20.48 \pm 0.67$ & $3.66 \pm 3.28$ & $20.80 \pm 0.88$ \\ \hline
   ${\rm NEa}_{+-}$ $40 M$ &  94.0 & $ -7.89 \pm 0.26$ & $1.04 \pm 0.08$ & $-18.92 \pm 0.63$ & $2.49 \pm 0.20$ & $19.08 \pm 0.63$ \\
                    $60 M$ & 114.0 & $ -7.89 \pm 0.21$ & $1.00 \pm 0.07$ & $-18.92 \pm 0.51$ & $2.40 \pm 0.17$ & $19.07 \pm 0.51$ \\ \hline
   ${\rm NEb}_{+-}$ $40 M$ &  72.0 & $-12.33 \pm 0.56$ & $2.03 \pm 0.64$ & $-29.41 \pm 1.34$ & $4.84 \pm 1.53$ & $29.81 \pm 1.35$ \\
                    $60 M$ &  92.0 & $-12.44 \pm 0.39$ & $2.00 \pm 0.55$ & $-29.67 \pm 0.93$ & $4.77 \pm 1.31$ & $30.05 \pm 0.94$ \\
  \hline \hline
 \end{tabular}
\end{table}
\end{center}

\begin{center}
\begin{table}[t]
  \caption{Basic parameters $M_f$ and $a_f$ of the post-merger hole,
  as calculated from radiated energy and angular momentum (``rad''),
  and from measured quasi-normal mode frequencies (``QNM''). All
  quantities based on extraction at $R_{\rm ext} = 60M$, except for
  $R_{\rm ext} = 30M$ for run ${\rm EQ}_{+0}$.}
\label{table:final_params}
 \begin{tabular}{c| c c c l c c| c c}
  \hline  \hline
   Run      & $M_{ADM}/M$ & $E_{rad}/M$ & $M_{f,rad}/M$ & $J_{ADM}/M^2$ & $J_{rad}/M^2$ & $(J/M^2)_{f,rad}$ & $M_{f,QNM}/M$ & $(a/M)_{f,QNM}$ \\
   \hline
  ${\rm NE}_{00}$  & 1.24 & $-6.14\times 10^{-4}$ & 1.24 & 0.0    &  0.0       & 0.0 & 1.2536 & 0.0252 \\
  ${\rm EQ}_{+0}$  & 1.00 & $-7.90\times 10^{-4}$ & 1.00 & 0.2    & $-1.101 \times 10^{-3}$ & 0.1992 & 1.0102 & 0.3507 \\
  ${\rm EQ}_{+-}$  & 1.00 & $-9.02\times 10^{-4}$ & 1.00 & 0.0    & $-0.000 \times 10^{-4}$ & 0.0& 1.0113 & 0.0198 \\
  ${\rm NE}_{+-}$  & 1.24 & $-8.19\times 10^{-4}$ & 1.24 & 0.0    & $-3.508 \times 10^{-4}$ & $2.33 \times 10^{-4}$ & 1.2585 & 0.0261 \\
  ${\rm NEa}_{+-}$ & 1.25 & $-8.16\times 10^{-4}$ & 1.25 & 0.0    & $-1.582 \times 10^{-4}$ & $1.01 \times 10^{-4}$ & 1.2676 & 0.0300 \\
  ${\rm NEb}_{+-}$ & 1.26 & $-1.10\times 10^{-3}$ & 1.26 & 0.2486 & $-5.000 \times 10^{-4}$ & 0.154 & 1.2582 & 0.2175 \\
  \hline
 \end{tabular}
\end{table}
\end{center}
\end{widetext}

\section{Discussion} 
\label{sec:discussion}

In this paper, we have addressed the roles of unequal masses and spins
in producing recoil kicks in head-on collisions. We have chosen sample
spins and mass ratios to explore how important spin effects are for
kicks, and how the effects of spin and mass ratio are related in
generating kicks.  We have carried out kick extractions from data sets
with mass ratios of 1 and $2/3$, and with spins on one or both holes.

We have observed that head-on collisions of spinning black holes can
produce significant kicks.  For the anti-aligned spin cases we have
studied, the spin-kicks are transverse to the direction of initial
separation. The magnitude of the kicks, and even the thrust curves
(Fig. \ref{thrust_Kidder_comp}), scale with the sum of the individual
black hole spin parameters, $a_1+a_2$.  These kicks easily exceed the
longitudinal kick produced in the case of unequal masses. To the
accuracy available in our simulations, we find that the kicks due to
the mass ratio and spins are indeed orthogonal and independent.  Both
the spin scaling and the orthogonality of spin-kicks and
mass-ratio-kicks are consistent with leading-order PN predictions for
these effects, as given by
Eqs.~(\ref{eqn:PN_PN_simp},\ref{eqn:PN_PSO_simp}).

It is remarkable that the PN predictions seem to describe our results.
The kick-producing radiation derived in our simulations is generated
in the systems' non-linear mergers and ringdowns, where the
assumptions behind the PN approximations clearly do not apply.  In
general terms, the PN analysis (see \cite{Kidder95a}) provides that
the unequal-mass-kick is produced by a coupling of the mass-quadrupole
and mass-octupole moments, while the spin-kick is produced by a
coupling of the mass-quadrupole and current-quadrupole moments.  At
least heuristically, we can consider these moments even in the
nonlinear problem.  In our problem, the presence of spin does not seem
to have a large effect on the mass moments, as is suggested by the
spin-independence of the mass-quadrupole-dominated $(l=2,m=2)$
waveforms, seen in Fig.\ref{EQ_WF}.  The leading effect of putting
spin on the black holes is to scale up the current-quadrupole moment.
For general spins, this term scales with $(\vec S_1/m_1-\vec
S_2/m_2)$.

In the more important case of inspiraling black holes, our results
suggest that the current-quadrupole effects should also provide strong
kicks, scaling in a similar way with spin. Because of the scaling with
$S/m$ we would generally expect the strongest spin-kicks for
nearly-equal-mass mergers. Our results support the use of PN
expressions (as in
\cite{Herrmann:2007ac,Baker:2007gi,Campanelli:2007ew}) to predict the
scaling of spin-kicks for black hole configurations that have not yet
been studied.

Since initial submission of this paper, several groups have
begun mode-based analysis of recoil kicks from binary inspiral
\cite{Schnittman:2007ij,Brugmann:2007zj}. Additionally, further investigation
of the maximum size and genericity of spin superkicks has been
undertaken
\cite{Campanelli:2007cg,Tichy:2007hk,Herrmann:2007ex,Lousto:2007db}.
These substantial predictions have opened the door to exciting new
possibilities in the astronomy of supermassive black holes and
galactic nuclei.

\acknowledgments

We would like to thank Sean McWilliams for useful discussions and
insight.

This work was supported in part by NASA grants O5-BEFS-05-0044 and
06-BEFS06-19.  The simulations were carried out using Project Columbia
at NASA Ames Research Center and at the NASA Center for Computational
Sciences at Goddard Space Flight Center.
D.C. was supported in part by The Korea Research Foundation and The
Korean Federation of Science and Technology Societies Grant funded by
Korea Government (MOEHRD, Basic Research Promotion Fund).
B.K. was supported by the Research Associateship Programs Office of
the National Research Council and the NASA Postdoctoral Program at the
Oak Ridge Associated Universities.

The {\tt PARAMESH} software used in this work was developed at the
NASA Goddard Space Flight Center and Drexel University under NASA's
HPCC and ESTO/CT projects and under grant NNG04GP79G from the
NASA/AISR project.


\end{document}